\documentclass[twocolumn]{aastex631}

\usepackage{multirow}

\newcommand{\magra}{\texttt{MAGRATHEA}}

\graphicspath{{./}{Figures/}}

\begin{document}

\title[Galactic Chemical Evolution and Exoplanets]{Effect of Galactic Chemical Evolution on Exoplanet Properties}

\author[0000-0003-2202-3847]{Jason H. Steffen}
\affiliation{Department of Physics and Astronomy, University of Nevada, Las Vegas, 4505 South Maryland Parkway, Las Vegas, NV 89154, USA}
\affiliation{Nevada Center for Astrophysics, University of Nevada, Las Vegas, 4505 South Maryland Parkway, Las Vegas, NV 89154, USA}

\author[0000-0003-4709-2689]{Cody Shakespeare}
\affiliation{Department of Physics and Astronomy, University of Nevada, Las Vegas, 4505 South Maryland Parkway, Las Vegas, NV 89154, USA}
\affiliation{Nevada Center for Astrophysics, University of Nevada, Las Vegas, 4505 South Maryland Parkway, Las Vegas, NV 89154, USA}

\author[0000-0002-0900-0192]{Robert Royer}
\affiliation{Department of Physics and Astronomy, University of Nevada, Las Vegas, 4505 South Maryland Parkway, Las Vegas, NV 89154, USA}
\affiliation{Nevada Center for Astrophysics, University of Nevada, Las Vegas, 4505 South Maryland Parkway, Las Vegas, NV 89154, USA}

\author[0000-0001-6009-8685]{David Rice}
\affiliation{Astrophysics Research Center (ARCO), Department of Natural Sciences, The Open University of Israel, Raanana 4353701, Israel}

\author[0000-0001-9504-3174]{Allona Vazan}
\affiliation{Astrophysics Research Center (ARCO), Department of Natural Sciences, The Open University of Israel, Raanana 4353701, Israel}

\begin{abstract}

We couple a simplified model for the galactic chemical evolution, with software that models the condensation of dust in protoplanetary disks and software that models the interior structure of planets in order to estimate the effects that the galactic chemical evolution has on the properties of planets as they form over time.  We find that the early abundance of elements formed from the evolution and death of high-mass stars (such as Oxygen, Silicon, and Magnesium) yields planets with larger mantles and smaller cores.  The later addition of elements produced in low-mass stars (such as Iron and Nickel) causes the planet cores to become relatively larger.  The result is planets that orbit older stars are less dense than planets orbiting younger stars.  These results are broadly consistent with recent observations of planet properties from stars of varying ages.

\end{abstract}


\section{Introduction} \label{sec:intro}

Beginning with the primordial mixture of light elements, the chemical composition of the galaxy evolves as nucleosynthesis processes occur in stars and their remnants.  Newly forming planets inherit the nucleosynthetic legacy of previous stellar generations. The processes that produce the various elements occur over different timescales.  Massive stars, with short lifetimes, enrich the Galaxy early with $\alpha$-elements such as Mg, Si, and O via core-collapse supernovae, whereas Fe-peak and other siderophile elements arise primarily from Type Ia supernovae and the deaths of low-mass stars, introducing a time delay of a few Gyr in their availability \citep{Kobayashi2020}.

The evolving chemistry of the galaxy causes the elemental ratios of planet-forming solids---and eventually planetary cores and mantles---change with time.  Indeed, the relative abundances of elements produced in low-mass and high-mass stars is known to correlate strongly with stellar age \citep{Nissen:2015,TucciMaia:2016}, and even the galactic orbital populations \citep{Ness:2019}.  Because Fe and other heavy siderophiles preferentially sink during planetary differentiation, the changing [$\alpha$/Fe] ratio, indeed the changing of the overall chemical abundances, in protoplanetary disks within the galaxy should leave a measurable imprint on the core-mass fraction (CMF), bulk density, and mantle properties of the forming planets (e.g., \citet{Cabral:2023}).

High-precision transit photometry and radial-velocity surveys have uncovered hundreds of small exoplanets with density estimates.  In addition, stellar age measurements, especially from asteroseismology, can have $<$10\% relative precision. Using 26 \textit{Kepler} planets, \citet{Weeks2025} reported a trend of increasing density with younger and younger planets.  (However, note that their linear extrapolation implies unphysical, 100\% iron planets at the current age of the Milky Way.)

Here we develop a framework that links a simple Galactic chemical-evolution (GCE) model informed by the Milky Way's star formation rate, dust-condensation chemistry, and an interior structure solver to estimate the relationship between stellar age and planet properties.  For this work we do not consider any potential spatial variation in the properties of stars or planets, as in \citet{Ness:2019,Teixeira:2025}.  We also leave for elsewhere the effects of total metallicity on the probability of forming planets that was seen in \citet{Boley:2024}.

In Sect.~\ref{sec:model} we outline the various elements of our modeling framework: the GCE model in Sect.~\ref{sec:stellar}, dust condensation model in Sect.~\ref{sec:disk}, and planet interior model in Sec.~\ref{sec:interior}.  We show how initial compositions, dust condensation, and planetary cores vary over the time since galaxy formation in Sect.~\ref{sec:result}. We discuss how these results relate to \citet{Weeks2025}, examine some of the details and limitations of this first application of our framework in Sect.~\ref{sec:discuss} and conclude in Sect.~\ref{sec:conc}.

\section{Model} \label{sec:model}

We use a straightforward model of the chemical evolution of the galaxy to explore its effects on the properties of forming planets.  Our model includes several components: the evolution of the stellar population and chemical enrichment of the galaxy, the chemical evolution of protoplanetary disks, and the structure and properties of planets that form from those disks.  For this work, we use the Sun and the solar system as a benchmark, with an implicit assumption that the Sun is representative of stellar composition at the time that it formed.  All other time-varying aspects of our model are measured relative to the Sun.

\subsection{Stellar Population Model} \label{sec:stellar}

Our model for the chemical enrichment of the galaxy begins with a simplified model for the evolution of the stellar population.  A more refined model of the stellar population is left for elsewhere.  In general, nucleosynthesis occurs through four primary mechanisms: Type II supernova, neutron star mergers, low-mass stellar evolution, and Type I supernova.  Here, we break the stellar initial mass function into two populations of stars: fast-evolving, high-mass stars; and slow-evolving, low-mass stars.  We use 12 solar masses as representative of high mass stars and 2 solar masses for low mass stars.

We combine the chemical elements that come from the supernova of high-mass stars and neutron star mergers into ``fast elements'' or ``fast process material.''  Similarly, those elements that come from exploding white dwarfs and dying low-mass stars are grouped together as ``slow elements'' or ``slow process material.''  We recognize that the coalescence of the orbits of neutron stars through the emission of gravitational radiation can take millions to billions of years, depending on the initial conditions of the pair.  Nevertheless, the first neutron star mergers will begin quickly after the first high-mass stars die and before many of the low mass stars make their contributions to the galactic chemistry\citep{Johnson2019}.

We also note that the timescale over which low-mass stars evolve occurs over billions of years, regardless of whether the material comes directly from the stellar evolution, or the subsequent white dwarf supernova arising from Roche lobe overflow from a nearby companion.  The elements formed from these processes are delayed by roughly two billion years from the time that the stars initially form---a lag similar to the lifetime of our fiducial, two solar mass stars.

From the star formation rate, which gives the number of stars produced over time, and the initial mass function, which gives the ratio of high and low-mass stars, we build a time series of relative abundances of fast and slow chemical elements beginning at a time 12 billion years in the past.  The mixture of fast and slow elements is normalized to the solar abundance when the Sun is born after eight billion years.  We then extend the normalized, relative abundances forward for an additional five billion years to today---producing a thirteen-billion-year time series.  We model the lifetimes of high mass stars as instantaneous since their lives are much shorter than the billion-year timesteps we consider.



Stepping through galactic time, we roughly follow the star formation rate (SFR) described in \citep{Snaith:2015,Haywood:2016,Maoz:2017}.  That is: 
\begin{enumerate}
\item A high SFR for 4 Gy \vspace{-0.1in}
\item A sharp drop of 93\% in the SFR for 2 Gy 
 \vspace{-0.1in}
\item A modest recovery to 26\% of the original rate until the present day  \vspace{-0.1in}
\item The integral of the SFR is normalized to unity.
\end{enumerate}
We use the Salpeter initial mass function (IMF) \citep{Salpeter:1955}:
\begin{equation}\label{eqn:SFR}
\xi (m) \sim 1/m^{2.35}
\end{equation}
to determine the relative frequencies of forming high mass (12 M$_\odot$) and low mass (2 M$_\odot$).  Since we are only considering relatively massive stars, deviations between the Salpeter model and more recent IMFs \citep[e.g.][]{Kroupa:2024} are negligible.  

While we approximate the high mass star lifetime as instantaneous, we use a 2 Gyr lifetime for our low mass stars, estimated using the mass-luminosity relationship:
\begin{equation}
\text{T}_\star = \left(10 \ \text{Gyr}\right) \ \left(\frac{M_\star}{M_\odot}\right) \left(\frac{1}{M_\star^{3.5}}\right).
\end{equation}
This age is roughly the same as the value from \citep{Salaris:2006} over this range in stellar masses.  The resulting material from the slow and fast processes as a fraction of the total material produced is shown in Figure \ref{fig:SFR} in the appendix.

\subsection{Galactic Chemical Evolution Model}


To find the relative elemental abundances to use in our protoplanetary disk models, we first need to convert our astonomical abundances (used to study stars) to cosmochemical abundances (used to study planets).  The cosmochemical abundance scale $N(El)$ is given by:
\begin{equation}
    N(El) = 10^{A(El)_0-1.61},
    \label{eq:CosmoScale}
\end{equation}
where $A(El)_0$ is the logarithmic astronomical abundance scale \citep{Lodders2003}.  We use the proto-Solar abundances from \citet{Lodders2003} to normalize all elements to $10^6$ silicon atoms.  Thus, $N(H)=2.431*10^{10}$ means $2.431*10^{10}$ hydrogen atoms for every $10^6$ silicon atoms.  

We assume at galactic time $t=0$, the stars comprise only primordial hydrogen and helium, so all stars have the solar ratio of H/He.  Since we only consider solids in our planet structure, neither H nor He contribute meaningfully to the condensates.  In addition, because the dust condensation software takes elements normalized to $10^6$ Si atoms, the abundances at $t=0$ (before any chemical enrichment occurs) are not useful.

The combination of the SFR and IMF produces a time series for the material produced by the fast process and slow process at each 0.5 Gyr time step.  The time series of fast and slow process material has arbitrary units, which we call $U_F$ and $U_S$, respectively.  These units are corrected to reproduce the Sun’s composition at t=8 Gyr.  

Given the total amount of metals produced by the fast and slow processes at each timestep ($i$), we calculate the non-normalized cosmochemical abundances, $N^*(El)$ using:
\begin{equation}
    N^*(El)_i=N^*(El)_{i-1}+\frac{\#A_{El}}{U_F}U_{F(i)}+\frac{\#A_{El}}{U_S}U_{S(i)},
\end{equation}
where the silicon-normalized abundances are calculated using:
\begin{equation}
    N(El)=N^*(El)\frac{N(\text{Si})}{N^*(\text{Si})}.
\end{equation}
For each timestep we normalize each element to $10^6$ Silicon atoms as indicated above.  This normalization has the effect of increasing the amount of hydrogen and helium at earlier times and reducing it at later times, which corresponds to small amounts of rocky material at early times, and more rocky material for stellar systems younger than the Sun, as expected.

Along with running our disk simulations using the nominal composition from each timestep in galactic history, we also consider systems that are enriched in high-mass material as well as low-mass material to examine the effects of local variations in these abundances.  \citet{Brewers2016} finds that C/O ratios vary between 0.2 and 0.7.  We use a range from C/O=0.333-0.75, as they both differ from the solar ratio (C/O=0.5) by a factor of 1.5.  We use observed variations in the C/O ratio at different ages because these elements largely represent the contributions from fast and slow material.  Oxygen comes almost entirely from high-mass stars and 75\% of Carbon comes from low-mass stars\citep{Johnson2019}.  Thus, we implicitly assume the observed C/O variation of 0.333-0.75 corresponds to stars formed at that time due to heterogeneous exposure to fast and slow materials.

For simplicity, we assume the maximum C/O ratio of 0.75 is due solely to slow process sources. Thus, we calculate 
\begin{equation}
    \frac{C}{U_{S}}=\left(\frac{C_\odot}{U_{S}}\right)\left(1 + \left(\frac{8}{9}\right) \left(\frac{\text{Slow}}{\text{Total}}\right)_{C} \right),
\end{equation}
or
\begin{equation}
    \frac{C}{U_{S}}=\left(\frac{C_\odot}{U_{S}}\right)\left(1 + \left(\frac{8}{9}\right) \left(0.75\right) \right),
\end{equation}
which yields C/O=0.75 at the time of the Sun's formation, t=8 Gyr.  The same calculation applies for all other elements with their own cosmochemical abundance $El_\odot$ and slow fraction $\frac{\text{Slow}}{\text{Total}}_{El}$.

To examine the effects of excess fast process material, we assume the minimum C/O ratio of 0.333 is due to increased oxygen from fast processes. 25\% of carbon comes from fast process material as well, thus, a minimum C/O of 0.333 requires a 56.1\% increase in fast process material:
\begin{equation}
    \frac{O}{U_F} = \left(\frac{O_\odot}{U_F}\right) \left(1 + 0.561 \left(\frac{\text{Fast}}{\text{{Total}}}\right)_{O}\right),
\end{equation}
thus,
\begin{equation}
    \frac{O}{U_F} = \left(\frac{O_\odot}{U_F}\right) \biggl( 1 + 0.561 \left(1.0\right) \biggr).
\end{equation}
We again normalize this enriched composition to the Sun's formation at 8 Gyr.

\begin{figure}
    \centering
    \includegraphics[width=0.45\textwidth]{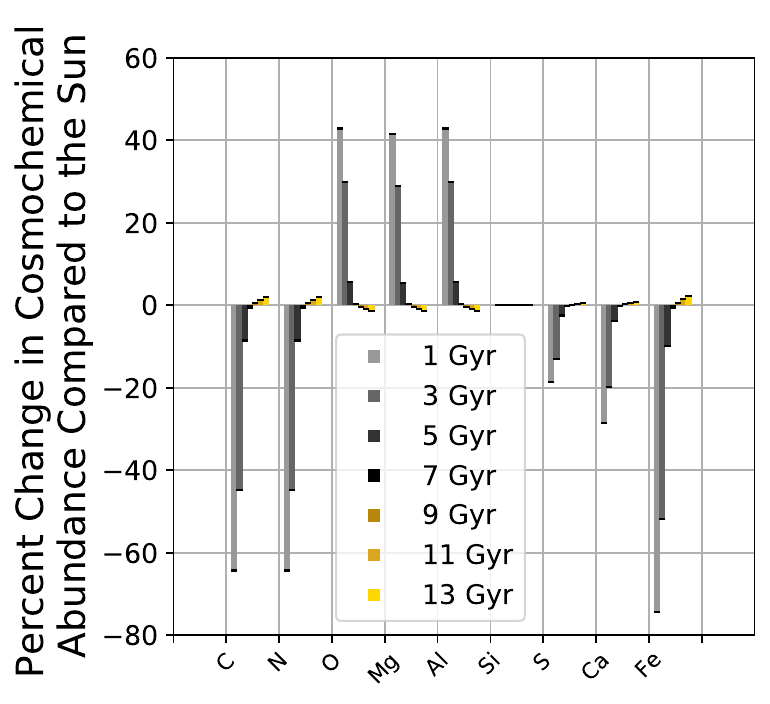}
    \caption{The percent change in initial cosmochemical abundance for select metals, normalized to Silicon and to the Solar abundances at $t=8$ Gyr, for each odd time step.}
    \label{fig:Ele}
\end{figure}

\subsection{Protoplanetary Disk Model} \label{sec:disk}

We now model how these changes in galactic composition impact the surface density and composition of condensed solids within protoplanetary disks.  We run chemical condensation models of the dust in the early stellar system with the partial dust condensation program developed in \citet{Li2020} and \citet{Shake2025}.  This code combines the disk evolution model in \citet{Cassen1996}, the dust advection model of \citet{Jacquet2012}, and the equilibrium chemical condensation code \texttt{GRAINS} \citep{Petaev2009}.  It was designed to reproduce the local refractory material in solar-like systems, and its decoupling from the disk, and does not model the condensation of volatile ices.

Following \citet{Li2020} and \citet{Shake2025}, we simulate the condensation of material over the first $\sim10^5$ years.  We consider disks around both 1 M$_\odot$ stars, which are common for stars with current exoplanet discoveries, and 2 M$_\odot$ stars, which are the progenitors for most polluted white dwarf observations.  For these simulations, we only consider condensation temperatures above 300K (we form rocks, but not ices).  We note that for the 2 M$_\odot$ stars we leave the disk parameters unchanged aside from the mass.  These parameters were originally optimized for solar-mass stars, implying that they are not optimized for higher masses, but we leave studies of the effects of differing disk parameters for elsewhere.

\subsection{Planetary Structure Model}\label{sec:interior}

Using the condensed material from our disk simulations, we then model the interior structure of planets that would form within our disks.  For the solar-mass model, we consider planets that form at five orbital radii: 0.5, 1, 2, 3, and 4 AU.  We assume the forming planets have two fully differentiated layers of core and mantle, and construct planets with masses ranging from 0.5 to 10 Earth-masses at each orbital distance.  We assume that all of the siderophile elements will be in the core (Mn, Fe, Co, Ni, Mo, Ru, Pd, W, Re, Os, Ir, Pt, and Au) \citep{goldschmidt}.  These account for 13 of our 33 tracked elements.  The remaining 20 elements are placed in the mantle.  

We initially considered using only iron in the core as well as considering all elements heavier than iron in the core.  Our choice of using siderophile elements increases the CMF over pure iron by 2.8\% at most. The difference in CMF between using all elements heavier than iron and using siderophile elements is at most 0.4\%.  Table \ref{tab:CMFs1sol} in the appendix contains the core mass fraction (CMF) and mantle mass fraction (MMF) values used at each radius and time for the one and two solar mass simulations.  Table \ref{tab:CMFs1solHL} shows the CMF and MMF values taken from a disk enriched in elements from either high or low mass stars.

We use the spherically symmetric interior structure model \magra\ \citep{Magrathea} to determine the final radii and interior structure of the planets\footnote{\magra\ can be accessed at \url{https://github.com/Huang-CL/Magrathea}}.  To probe the relationship between total mass, CMF/MMF, and radius, we fix the surface temperature to be 300 K, and limit our simulations to planets without an atmosphere or hydrosphere.  Volatile layers would add additional degeneracies, but even the presence of some volatiles should still have a changing CMF with time \citep{Rogers:2025}, though a large volatile fraction would likely affect the core and mantle composition, which would affect the CMF in a different manner than what we outline here \citep{Luo:2024}.  We use the default material properties and phase transitions as in \citet{Rice2025} and an adiabatic temperature gradient without discontinuities throughout the planet.

\section{Results} \label{sec:result}

The details of the solid surface density of elements in the disk after chemical condensation are shown in the Appendix, Figures \ref{fig:Disk1M} and \ref{fig:Disk2M} for a $1M_\odot$ star and $2M_\odot$ star respectively.  The primary difference between the results of the models is that the $2M_\odot$ star does not have significant condensation of material inside of $\sim0.5$AU, which may change with a more thorough examination of how stellar mass effects disk parameters.  With our model, the compositional difference caused by stars of different masses is small.  

Figure \ref{fig:Ele} shows the initial abundances of several important chemical elements over galactic time normalized to the Solar abundance.  A similar, but more comprehensive graph is shown in the Appendix.  Elements from slow processes show smaller values early in the galaxy's history while elements from fast processes show the opposite behavior. All elements heavier than H and He increase in an absolute sense over time, however since our results are normalized to Silicon, it remains constant and the overall increase in metals is not seen directly in this figure.

At 1 Gyr, we find Iron levels in the condensed materials are below Silicon and Magnesium.  By 3 Gyr, the Iron abundance nearly equals Magnesium and Silicon by mass.  By 8 Gyr, and through 13 Gyr, the Iron abundance nearly equals Oxygen by mass.  The growth in Iron abundance suggests that cores will become more prominent as time goes on. 
Similar to the trend with Iron, Calcium levels are below Aluminum until 8 Gyr where they are nearly equal, and Carbon condensation is low at all ages, as the initial C/O ratio only reaches $\sim$0.52 by 13 Gyr.  This result implies that the mantle composition of most planets will be similar to the Earth.

In general, the rate of change of the elemental ratios slows dramatically after 6 Gyr due to the slowing of the star formation rate.  \citet{Bond2010} and \citet{Shake2025} showed that carbon does not become a primary constituent of planet composition until the C/O ratio becomes larger than $\sim0.8$. Nevertheless, since 75\% of carbon comes from low-mass stars while 100\% of oxygen comes from high-mass stars, the C/O ratio of stars will continue to slowly increase into the future as smaller and smaller stars complete their lifecycles.

\begin{figure*}[!ht]
    \includegraphics[width=0.48\textwidth]{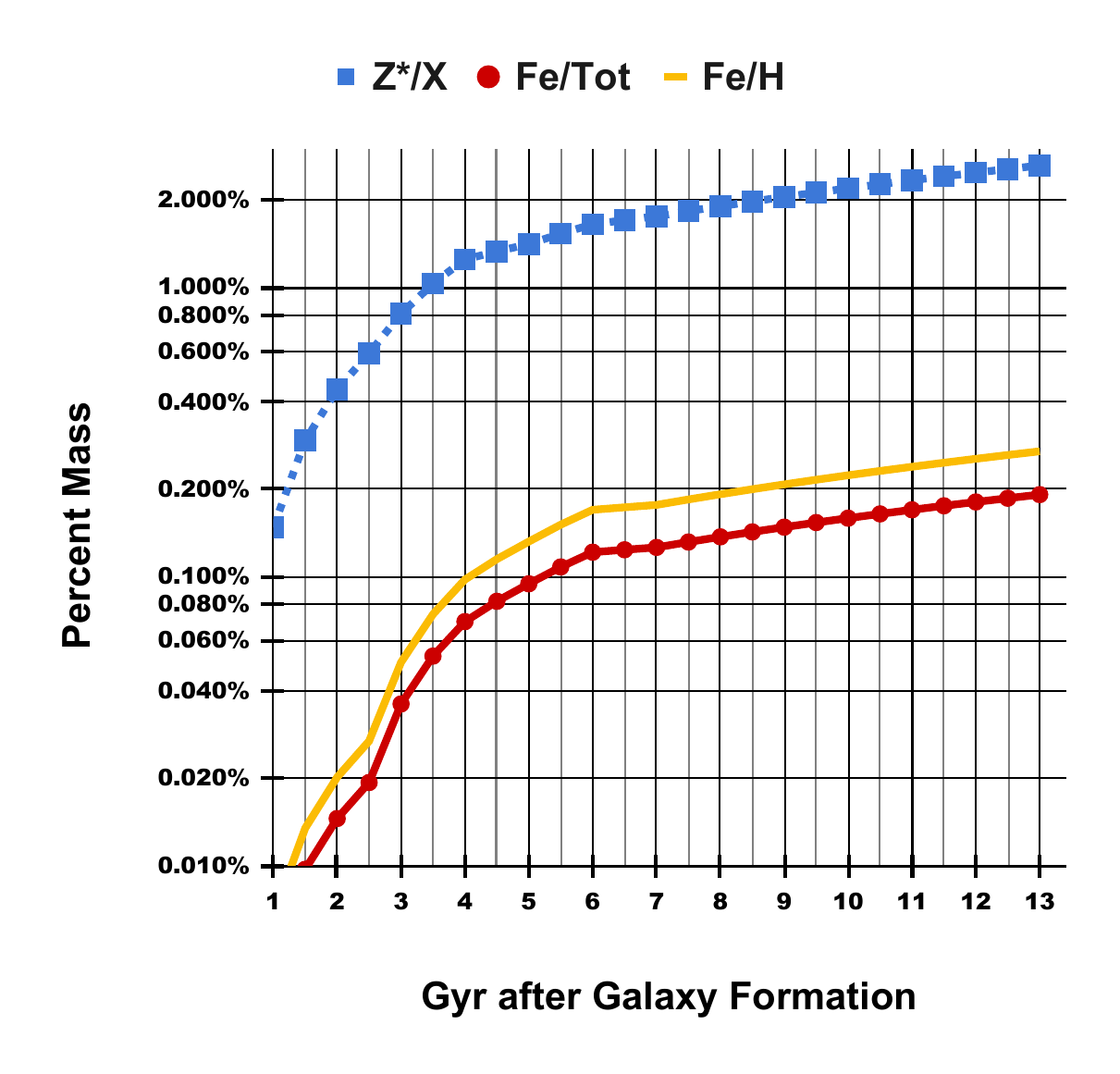}
    \includegraphics[width=0.48\textwidth]{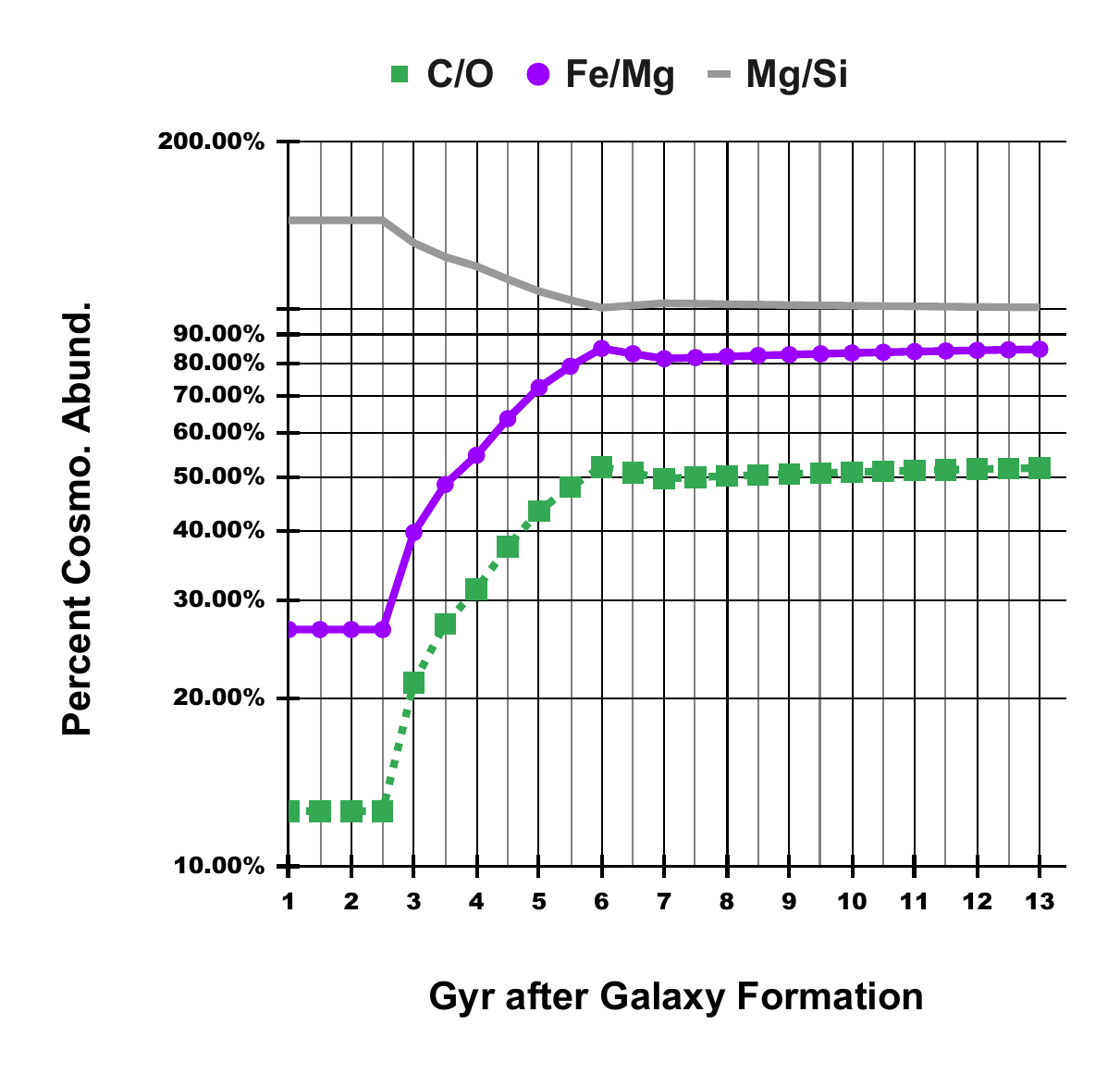}
    \caption{(Left) The initial percent mass of $Z_0^*/X_0$, iron over total elements, and iron over hydrogen. $Z_0^*$ is the sum of all metals used in our dust condensation program. The exclusion of the other metals results in our Solar $Z_0^*/X_0=1.9\%$ compared with \citet{Lodders2003}'s $Z_0/X_0=2.1\%$. (Right) The initial percent cosmological abundance of select elemental ratios.}
    \label{fig:Ratiosfg}
\end{figure*}

Figure \ref{fig:Ratiosfg} shows the change in the initial chemical composition through important elemental ratios.  The left of Figure \ref{fig:Ratiosfg} shows metallicity and iron fractions by mass. This Fe/H abundance agrees with several SFR profiles tested in \citet{Kobayashi2020}. The right of Figure \ref{fig:Ratiosfg} shows elemental ratios by percentage of the cosmochemical abundance ratios. The C/O ratio is useful for comparing the relative abundances of fast and slow process materials.  The Fe/Mg ratio is useful as a trace for the expected mass ratio of the core and the mantle.  The Mg/Si ratio determines whether volcanic rocks would be felsic, like granite, or mafic, like basalt.  This ratio is important since \citet{Bond2010} predict that silicon-depleted planets will have thick crusts that may prevent extrusive volcanism and tectonic activity. Figure \ref{fig:EleApp}, in the appendix, shows the fractional change in the initial cosmochemical abundance (relative to the Sun) of all elements included in our dust condensation program.

\begin{figure}
    \includegraphics[width=0.48\textwidth]{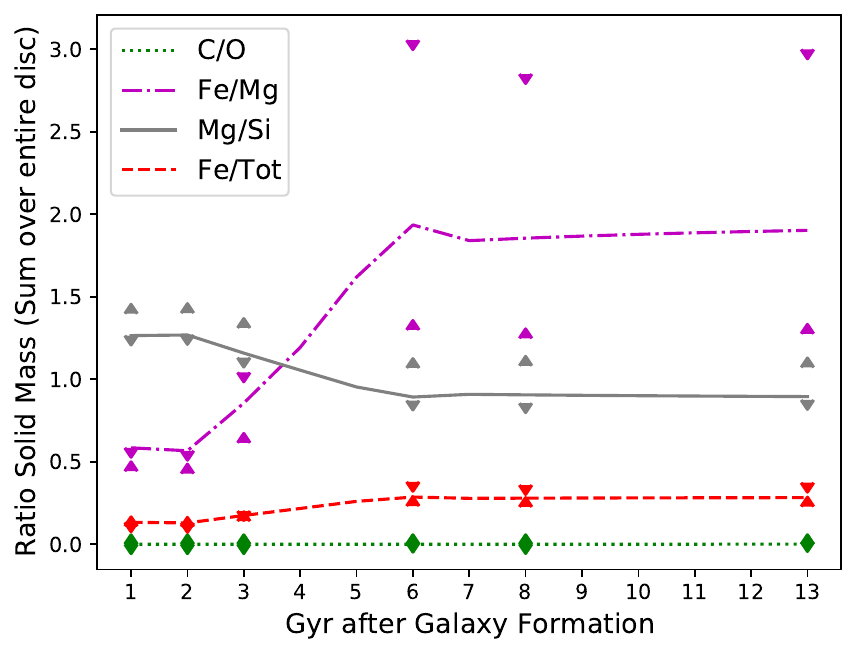}
    \caption{The elemental ratio of solid material after chemical condensation around a $1M_\odot$ star of varying age. The elemental ratio represents the total sum of those elements over the entire disc. Some areas of the disc can experience significant differences from the ratios shown. Discs around $2M_\odot$ stars show small differences that are imperceptible. Upward triangles represent the corresponding fast-material enriched simulations. Downward triangles represent the slow-material enriched simulations. The base of each triangle indicates enriched material ratios.}
    \label{fig:Ratio1M}
\end{figure}

Figure \ref{fig:Ratio1M} shows select elemental ratios calculated by summing the mass of each element's solid material over the entire disc. These elements condense as molecular species in our model, but while molecular species can change from chemical processes, the bulk elements will remain largely the same. The drawback for showing elemental ratios with the method in Figure \ref{fig:Ratio1M} is that some areas of the disc can have much higher or lower levels of different elements.  Elemental results at each location in the disc are shown in Figures \ref{fig:Disk1M}-\ref{fig:Disk2M} in the appendix.

We mark the results of our enriched material simulations with triangular markers; when two markers form a diamond, there is little to no change from the nominal model.  The highest initial C/O ratio simulated is 0.78 at 13 Gyr, which results in $\sim1.3$ percent total solid mass carbon over the entire disc.  This fact implies that carbon rich planets are more likely to exist in younger or future systems.  The behavior of the final, condensed elemental ratios closely matches the initial ratios from Figure \ref{fig:Ratiosfg}.

\begin{figure*}
    \includegraphics[width=0.48\textwidth]{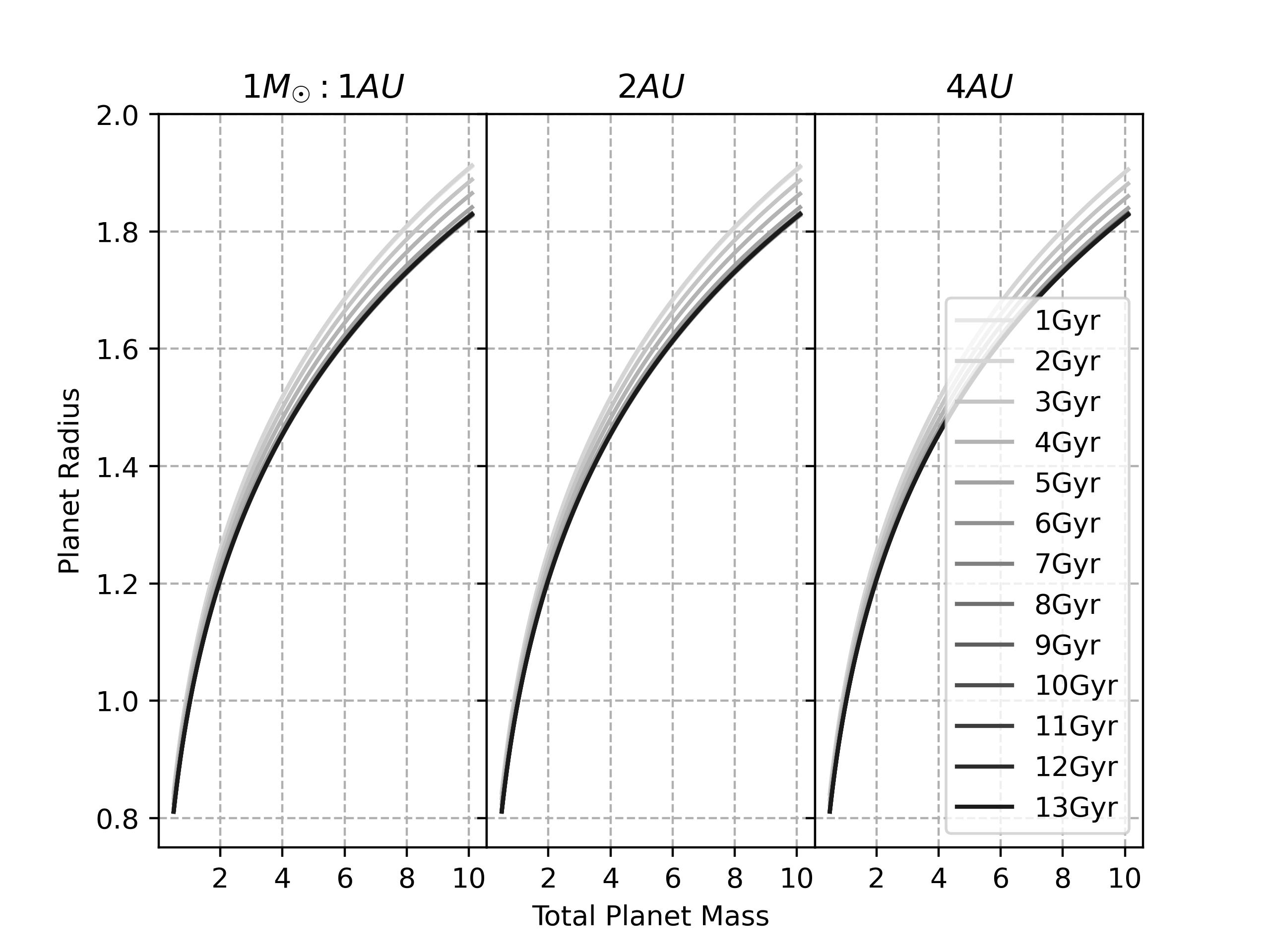}
    \includegraphics[width=0.48\textwidth]{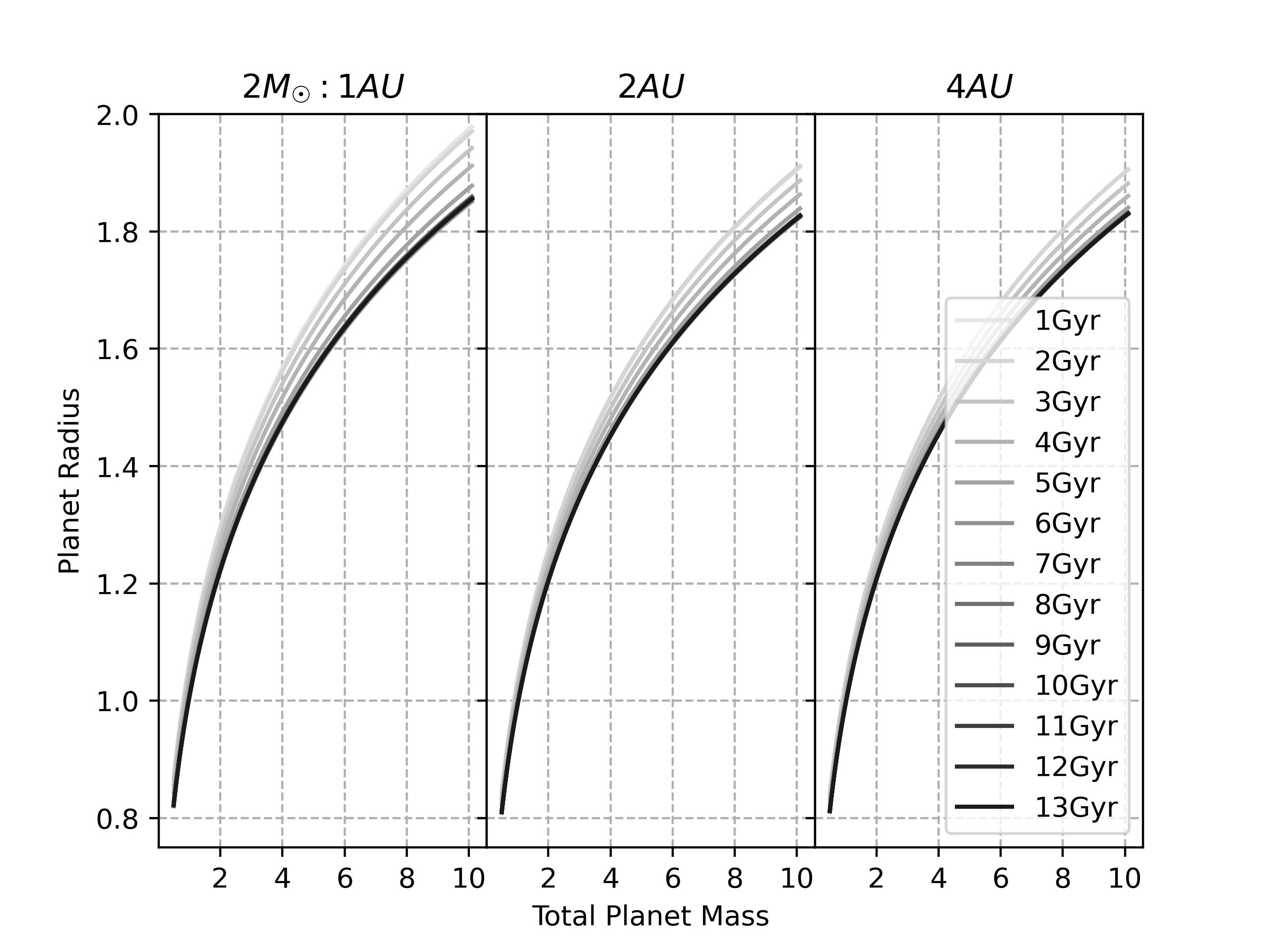}
    \caption{Radius vs mass for planets formed at different times and selected orbital radii. Color indicates the time at which the planet was formed, from 1 to 13 Gyr.  The left panel shows three orbital radii around a 1 Solar mass star.  The right panel shows the same orbital radii around a 2 Solar mass star.}
    \label{fig:MRfigs}
\end{figure*}

Ultimately, the bulk planet density increases in time along with the fraction of heavy elements.  Figure \ref{fig:MRfigs}, shows the radius of planets built from material that condenses at three different distances from the two fiducial host stars. The radii of the planets decrease with time (i.e., the density increases with time).  The largest changes in planet density occurring between two and five billion years. During this interval, planet radii at all distances decrease between 1.1\% and 1.3\% compared to planets formed later in galactic time, with planet sizes changing most dramatically for planets orbiting closest to their host stars.  In the case of planets orbiting 0.5 AU from the one solar mass star, the changes are 1.2\% to 1.8\%.  We see similar changes in planet radius for planets orbiting a two solar mass star at 1 AU.

\section{Discussion} \label{sec:discuss}

\begin{figure}
\includegraphics[width=0.49\textwidth]{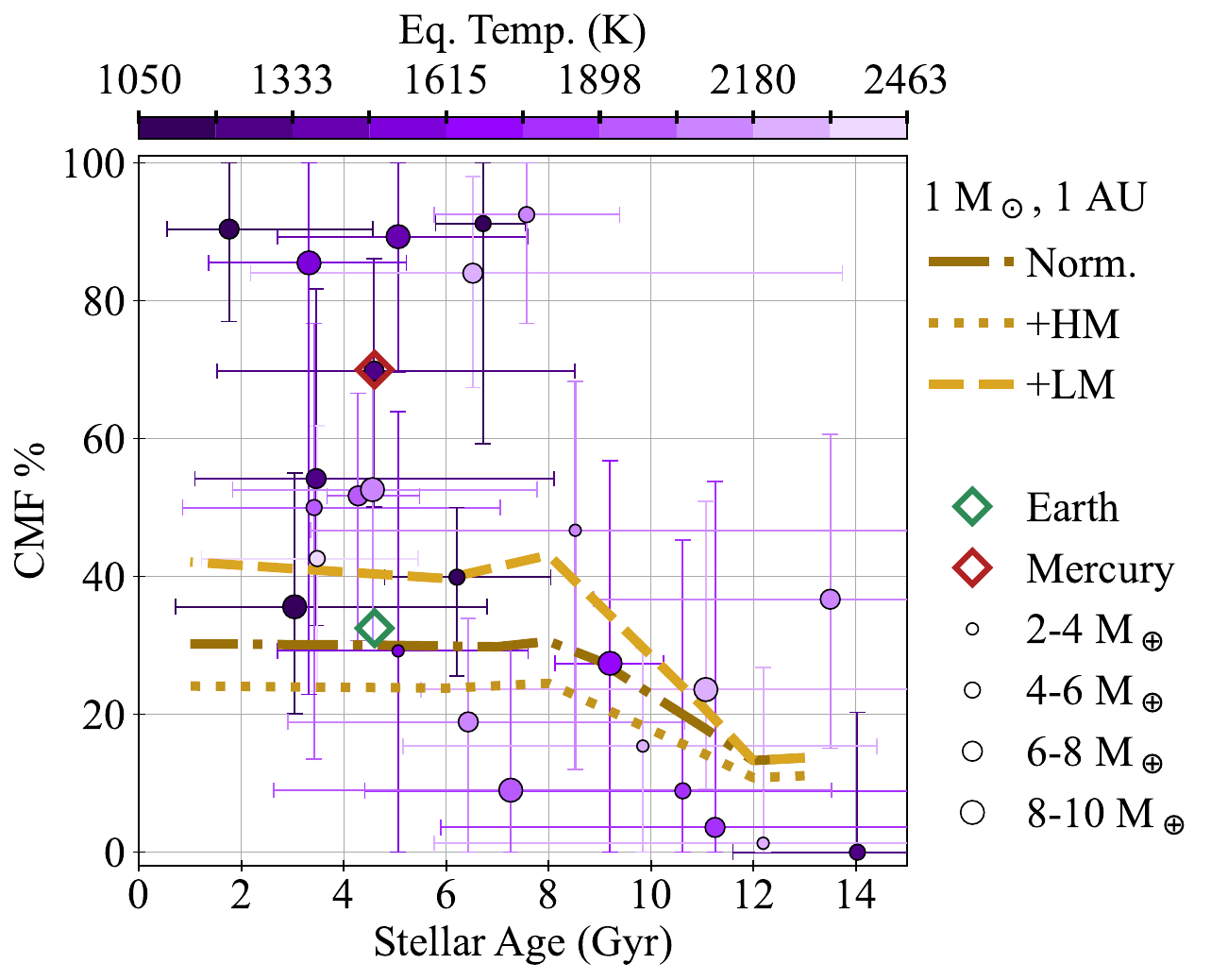}
\caption{The stellar age and CMF (\%) of planets from \citet{Weeks2025} with 1$\sigma$ uncertainties. The CMF (\%) is found with \magra\ given the masses and radii in \citet{Weeks2025}. The planet's equilibrium temperature is represented by the color and the planet's mass by the size of the circle. Three gold lines of CMF with age from our results are shown measured at 1 AU around a solar-mass star (note that this distance is exterior to most detected exoplanets, thus there is an implicit assumption that the observed planets migrated inwards after formation).  ``Norm.'' is when elemental abundances are normalized to the solar system while ``+HM'' is enriched in high-mass material and ``+LM'' has increased low-mass material. Earth and Mercury also shown for reference. Note that the x-axis in this case is age of the star and thus inverse of the previous plots.}
\label{fig:weeksplanets}
\end{figure}

Planets that form around early generation stars will have lower iron content, implying smaller cores as observed in \citet{Weeks2025}.  We estimate the CMF of each of the 26 planets from \citet{Weeks2025} using our interior model described in Sect.~\ref{sec:interior} and a composition finder developed in \citet{Rice2025}.  Our results are shown in Figure \ref{fig:weeksplanets}.  To get these results, for each planet we draw 100 samples of mass and radius from two Gaussians with mean and standard deviation set by the mass, radius, and uncertainties reported in \citet{Weeks2025}, and a surface temperature equal to the null albedo effective temperature of the planet.

We then use a two-layer model of mantle and core to find the CMF needed to match the observed radii in our sample to within 0.05\%.  Realizations that are too dense to be modeled with 100\% core are treated as 100\% CMF and those that are less dense than a 100\% mantle planet are treated as 0\% CMF. The 16th, 50th, and 84th percentiles of CMF verses the stellar age are shown in Figure \ref{fig:weeksplanets}.  This figure includes results from our nominal model as well as our models that are enhanced in fast and slow process materials.

Our results show a shallower relationship for CMF verses stellar age than the linear fit from \citet{Weeks2025}, with our CMF increasing to roughly 30-40\% for planets in young systems.  Only two planets around a host stars with an age less than 8 Gyr have a median CMF under 20\%. In the oldest systems, five of nine planets have a median CMF under 20\%. Our results agree with the overall trend of younger planets being more dense due to the influx of slow-process materials.

Our model does not predict the large population of super-Mercuries with a CMF above 60\% that the analysis from \citet{Weeks2025} suggests.  The existence of a significant population of super-Mercuries is a subject of debate.  In one case, a number of super-Mercury candidates in \citet{Adibekyan2021} were found to have lower densities after subsequent observations in \citet{Brinkman2024}. The formation of high CMF planets may be a result of the separation of iron in the protoplanetary disk by magnetic fields in the inner parts of the disk \citep{Bogdan2023}, or from giant collisions stripping away the mantle \citep{Marcus2010} (though collisions that aren't completely erosive will tend to drive planet composition to the median composition of their neighborhoods \citep{Ferich:2025}).


Our results suggest that early planets have higher Mg/Si ratios, which make Si-depleted planets with thick crusts, as predicted by \citet{Bond2010}. Note that the planets they consider occur around stars with Mg/Si ratios of 1.91 while our models center near a more modest ratio of 1.4.  Another interesting trend is the calcium-to-aluminum ratio. CAI's are a relatively small component of material in disc material of any age. However, the ratio of calcium-to-aluminum in these CAIs will vary over time, with implications for potential measurements of material from interstellar objects. If CAIs are found in interstellar material, their Ca/Al ratio may be useful in determining the properties of the star they came from.

Another, more subtle trend is the relationship between elements of similar sources.  For instance, aluminum and oxygen are both purely sourced from high-mass material in our model, and magnesium is nearly so.  Meanwhile, iron comes predominantly from low-mass sources, while carbon and calcium are sourced from both high-mass and low-mass stars. Our model suggests that material with high Ca/Al ratios will also have high Fe/Mg and C/O ratios, and vice versa.  Understanding how these different elements relate to one another in terms of their nucleosynthesis sources can help in making reliable estimates of elemental abundances when stellar spectra data is unreliable or nonexistent for some elements.


There are a number of ways we anticipate improving our model, including: using a more sophisticated star formation rate, using a complete initial mass function (which may be time dependent), separating out all of the different nucleosynthesis sources where each is given a separate time evolution instead of using only two timescales, and using a different disk evolution model---especially for stars of differing masses.  Another area that could have an effect on planet properties is the compositional variation that occurs during the formation process due to collisions among planetesimals (e.g. \citet{Marcus2010,Ferich:2025}).

\section{Conclusions} \label{sec:conc}

We constructed a physics-based framework that links a galactic chemical-evolution model to disk condensation chemistry and a planetary interior solver to predict the changes in the properties of exoplanets that form over time.  We looked specifically at the core mass fraction, bulk density, and mantle mineralogy.  In general, we find the following:

\begin{itemize}
    \item Planets become more dense as the galaxy evolves due to the increase in elements formed in low mass stars, as seen in the observations from \citep{Weeks2025}.  This means that the core mass fraction grows and the radii of planets decrease over galactic time.
    \item Early, between 2 to 6 Gyr, our model shows a rapid increase in Fe/Mg and C/O ratios while there is a decrease in Mg/Si.
    \item The mass fraction of Fe in solids in protoplanetary disks doubles between 1 Gyr to 5 Gyr after galaxy formation while remaining nearly constant later in the galaxy's evolution.
    \item Enrichment from different sources of stellar nucleosynthesis changes the core mass fraction of planets that formed at the same time by more than 10\%.
\end{itemize}

Upcoming PLATO asteroseismic ages \citep{Rauer:2025}, coupled with precise densities and JWST-era stellar abundance catalogs, will provide the leverage to measure the CMF trend predicted here. The framework presented can be extended with a full initial-mass-function GCE model, disk evolution tuned to stellar mass, stochastic enrichment events, and dynamical collisional evolution. Such extensions will establish a more accurate understanding of the changing properties of planets over time and turn rocky planets into probes of the Galaxy’s history.







\begin{acknowledgments}
We acknowledge support from the Graduate College and the Nevada Center for Astrophysics at the University of Nevada, Las Vegas (UNLV), the National Supercomputing Institute and Cherry Creek computing cluster at UNLV, as well as NASA under grant 80NSSC23M0104. AV acknowledge support by ISF grants 770/21 and 773/21.
\end{acknowledgments}

\vspace{5mm}
\facilities{UNLV, Cherry Creek}

\software{\textsc{Magrathea} \citep{Magrathea},
    \textsc{NumPy} \citep{numpy}, 
    \textsc{Matplotlib} \citep{matplotlib}
}

\clearpage

\appendix

\section{Appendix}

\restartappendixnumbering

We include in this appendix several tables of our simulation conditions and results, as well as further information about our methods.

\subsection{Notes on Chemical Abundances}
Our chemical condensation model and our galactic chemical evolution model use the cosmochemical abundance scale, $N(El)$. The cosmochemical abundance scale is common in the study of meteorite samples due to its abundance values being linear and often normalized to silicon. The astronomical abundance scale, $A(El)$, has logarithmic abundance values and is common in the study of stellar abundances and spectra. Equation \ref{eq:CosmoScale}, from \citet{Lodders2003}, can be used to convert between the two scales.

\begin{figure*}
        \centering
    \includegraphics[width=0.9\textwidth]{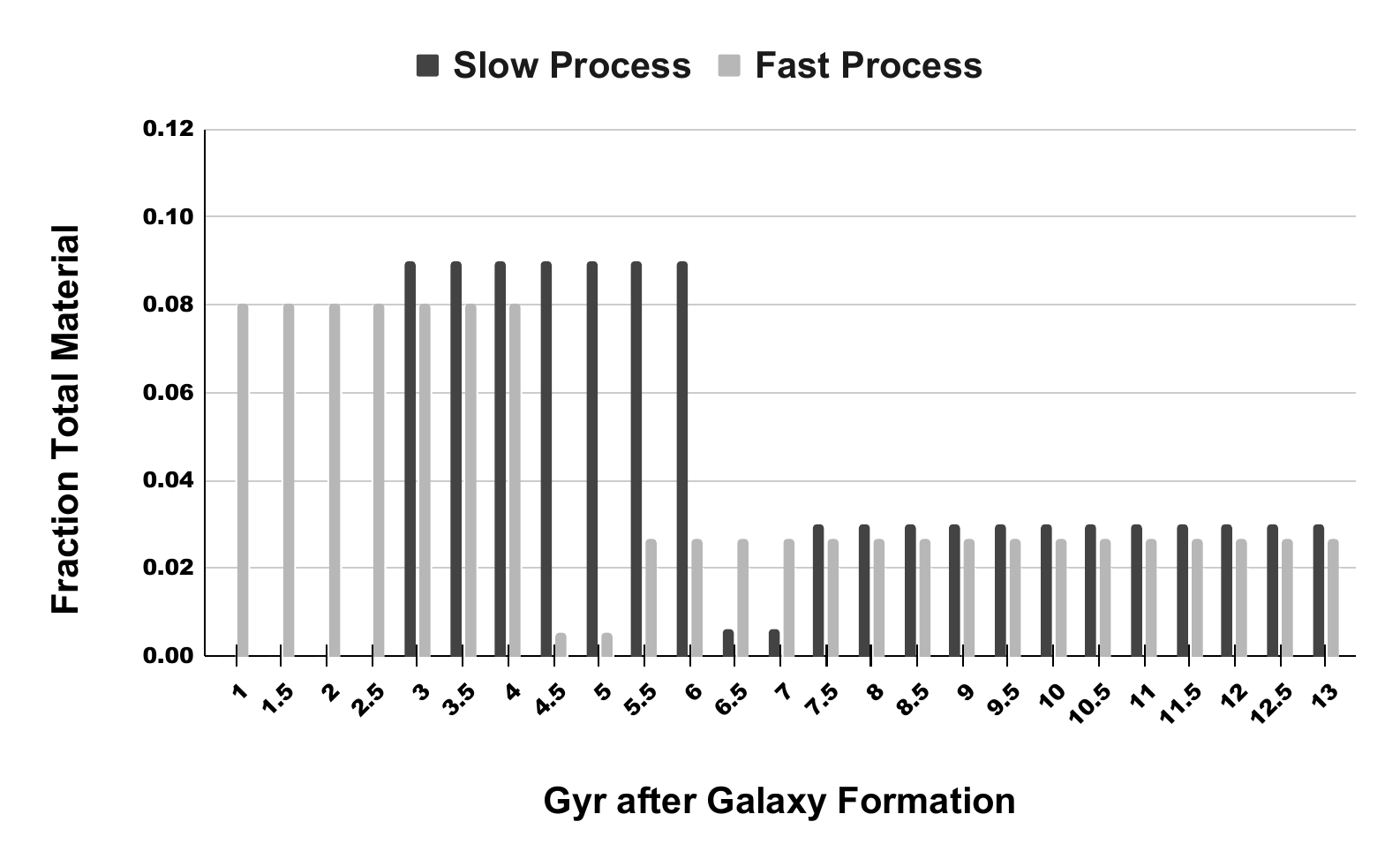}
    \caption{The amount of normalized material produced through fast processes and slow processes at each time step using the empirical star formation rate from \citep{Snaith:2015} and initial mass function from Equation \ref{eqn:SFR}.}
    \label{fig:SFR}
\end{figure*}

Figure \ref{fig:SFR} shows the amount of material produced by fast and slow processes from the empirical star formation rate we use from \citep{Snaith:2015} and our initial mass function \citep{Salpeter:1955}. Table \ref{tab:sims} shows the fraction of each element, $\frac{Fast}{Total}$, that is produced by fast or slow processes and the corresponding number of atoms of each element produced per unit of material, $\frac{\#A_{El}}{U_F}$. 

\begin{table*}[]
	\caption{The fraction of each element that we assume comes from fast or slow processes from \citet{Johnson2019}. Hydrogen and Helium are not included. There are $\sim0.048$ units of fast synthesized material (i.e. $U_F$) and $\sim0.46$ units of slow synthesized material (i.e. $U_S$) when the Sun forms at $t=8$ Gyr.}
	\label{tab:sims}
	\centering
        \scriptsize
        \tabcolsep=0.07cm
    \begin{tabular}{lcccccccccccccccccccc}
		\hline
		\hline
		\multicolumn{1}{c|}{\textbf{Elem}} & C & N & O & Na & Mg & Al & Si & P & S & Cl & K & Ca & Ti & Cr & Mn & Fe & Co & Ni & Cu & Ga \\
            \hline
		\multicolumn{1}{c|}{$\frac{Fast}{Total}$} & 0.25 & 0.25 & 1 & 1 & 0.99 & 1 & 0.7 & 0.97 & 0.57 & 0.83 & 0.81 & 0.5 & 0.34 & 0.24 & 0.18 & 0.32 & 0.32 & 0.29 & 0.42 & 1 \\
        \multicolumn{1}{c|}{$\frac{Slow}{Total}$} & 0.75 & 0.75 & 0 & 0 & 0.01 & 0 & 0.3 & 0.03 & 0.43 & 0.17 & 0.19 & 0.5 & 0.66 & 0.76 & 0.82 & 0.68 & 0.68 & 0.71 & 0.58 & 0 \\
            \hline
		\multicolumn{1}{c|}{$\frac{\#A_{El}}{U_F}$} & 3.7e11 & 1.0e11 & 3.0e12 & 1.2e10 & 2.1e11 & 1.8e10 & 1.5e11 & 1.7e9 & 5.3e10 & 9.1e8 & 6.3e8 & 6.6e9 & 1.7e8 & 6.5e8 & 3.5e8 & 5.6e10 & 1.6e8 & 2.9e9 & 4.6e7 & 7.5e6 \\
        \multicolumn{1}{c|}{$\frac{\#A_{El}}{U_S}$} & 1.2e11 & 3.2e10 & 0 & 0 & 2.2e8 & 0 & 6.5e9 & 5.5e6 & 4.2e9 & 1.9e7 & 1.5e7 & 6.9e8 & 3.5e7 & 2.1e8 & 1.6e8 & 1.2e10 & 3.4e7 & 7.4e8 & 6.7e6 & 0 \\
		\hline
		\hline
	\end{tabular}
    \begin{tabular}{lccccccccccc}
		\hline
		\hline
		\multicolumn{1}{c|}{\textbf{Elem}}& Ge & Mo & Ru & Pd & Hf & W & Re & Os & Ir & Pt & Au\\
            \hline
		\multicolumn{1}{c|}{\textbf{$\frac{Fast}{Total}$}}& 1 & 0.38 & 0.67 & 0.54 & 0.44 & 0.44 & 0.91 & 0.91 & 0.99 & 0.95 & 0.94\\
        \multicolumn{1}{c|}{\textbf{$\frac{Slow}{Total}$}}& 0 & 0.62 & 0.33 & 0.46 & 0.56 & 0.56 & 0.09 & 0.09 & 0.01 & 0.05 & 0.06\\
            \hline
		\multicolumn{1}{c|}{$\frac{\#A_{El}}{U_F}$}& 2.5e7 & 2.1e5 & 2.7e5 & 1.6e5 & 1.6e4 & 1.2e4 & 1.0e3 & 1.3e5 & 1.3e5 & 2.7e5 & 3.8e4\\
        \multicolumn{1}{c|}{$\frac{\#A_{El}}{U_S}$}& 0 & 3.5e4 & 1.4e4 & 1.4e4 & 2.1e3 & 1.6e3 & 1.0e2 & 1.3e3 & 1.4e2 & 1.5e3 & 2.6e2\\
		\hline
		\hline
	\end{tabular}
\end{table*}

\begin{figure*}
    \centering
    \includegraphics[width=0.96\textwidth]{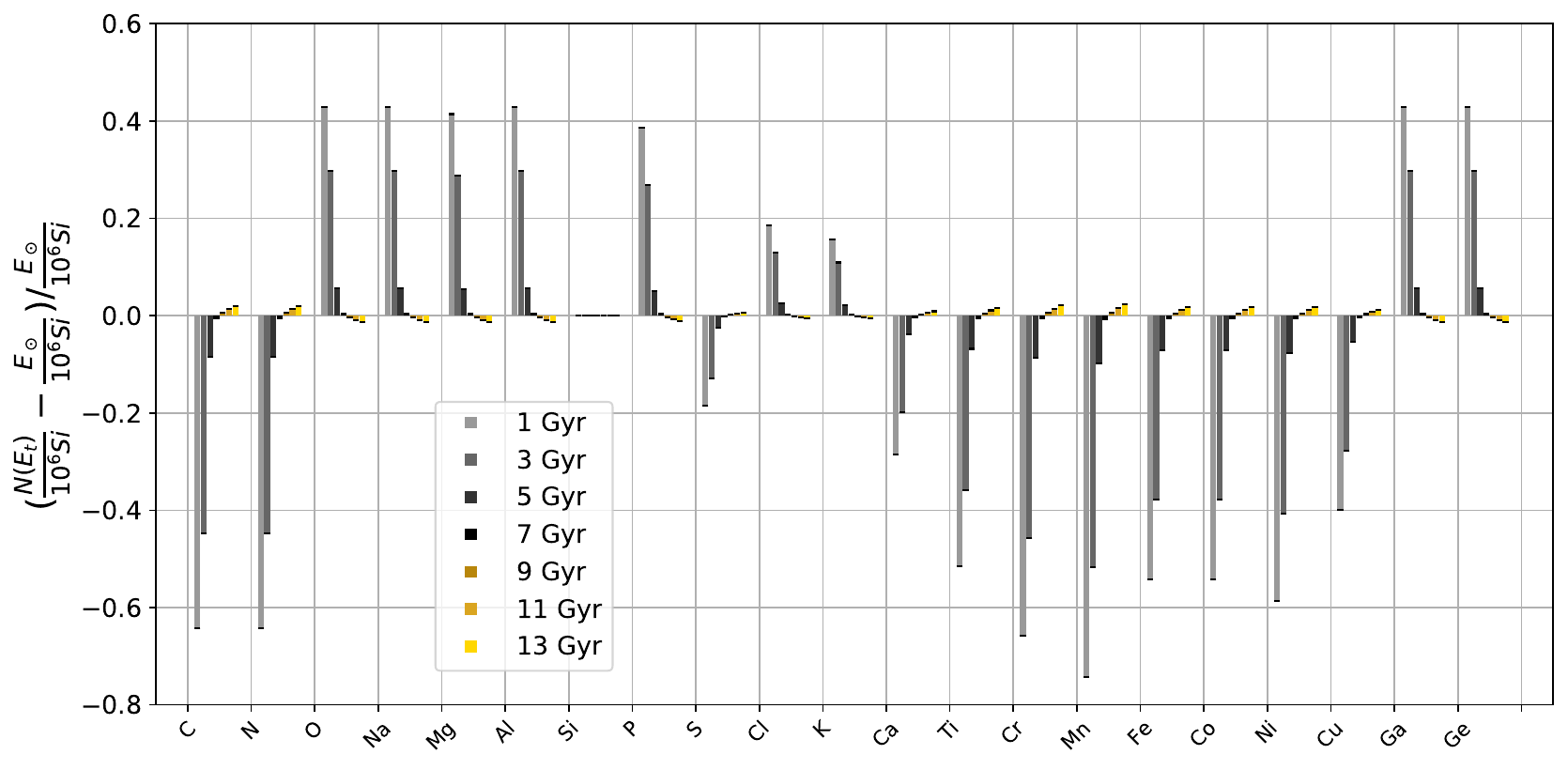}
    \includegraphics[width=0.65\textwidth]{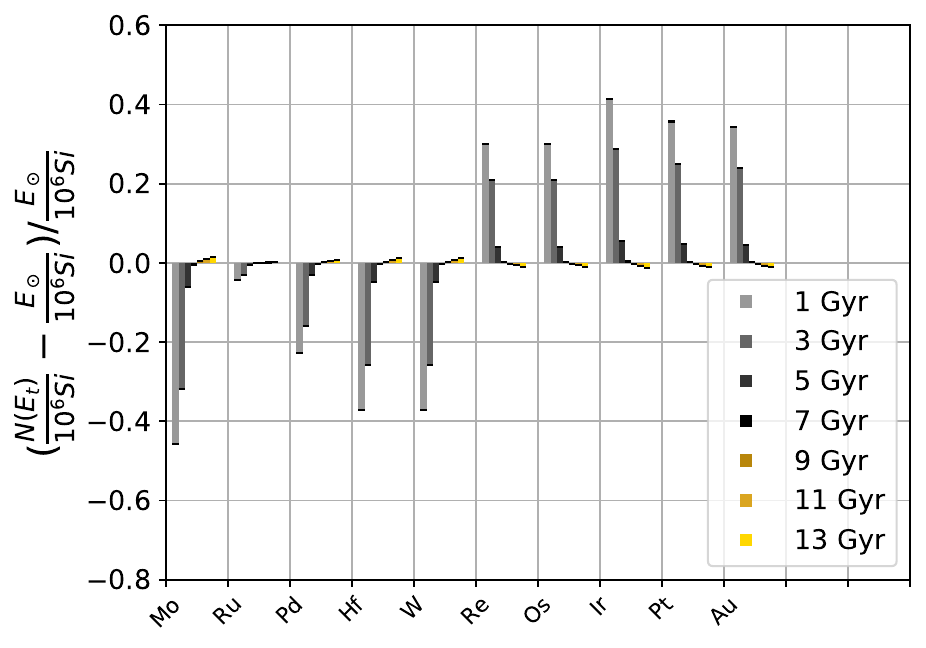}
    \caption{The fractional change in initial cosmochemical abundance of all metals included in dust condensation, normalized to Solar at $t=8$ Gyr, for each odd time step. (Top) C-Ge (Bottom) Mo-Au}
    \label{fig:EleApp}
\end{figure*}

Figure \ref{fig:EleApp} shows the changes in relative abundance over time of all elements that we consider in our simulations.

\begin{figure*}
    \includegraphics[width=0.96\textwidth]{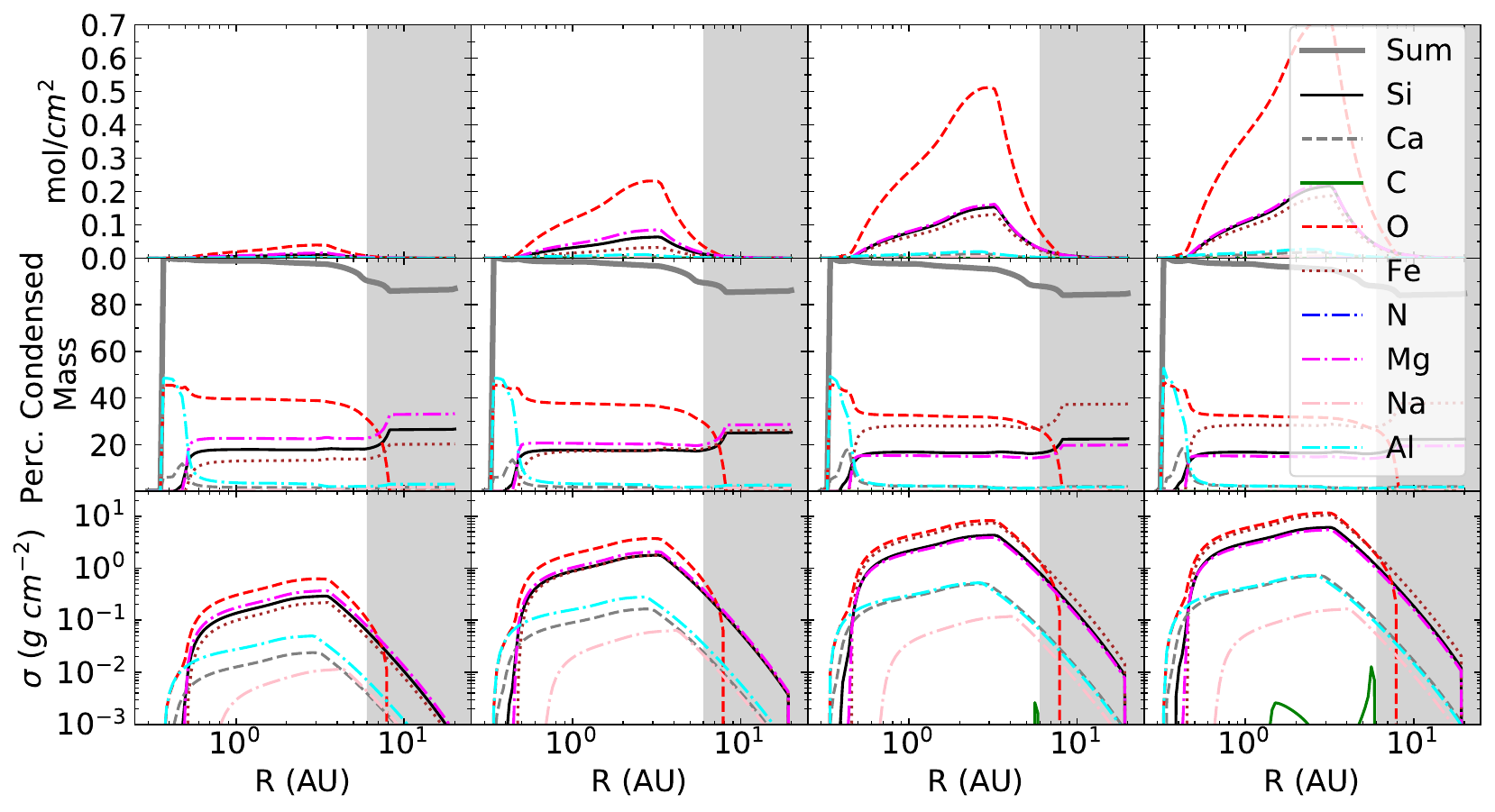}
    \caption{The final solid surface density of various elements in disks around 1 M$_\odot$ stars. Select elements are shown using three different units and with four different stellar abundances. From left-to-right columns, the results of a star formed with the elements present at 1 Gyr, 3 Gyr, 8 Gyr, and 13 Gyr. (Top row) Surface density on a linear y-axis with units of moles per square centimeter. (Middle row) Percent solid mass of each element's solid phase (i.e. $\frac{Oxygen_{Solid}}{Total_{Solid}}*100$). The 'Sum' line is the sum of all elements shown in the plot. (Bottom row) Surface density on a logarithmic y-axis with units of grams per square centimeter.}
    \label{fig:Disk1M}
\end{figure*}

\begin{figure*}
    \includegraphics[width=0.96\textwidth]{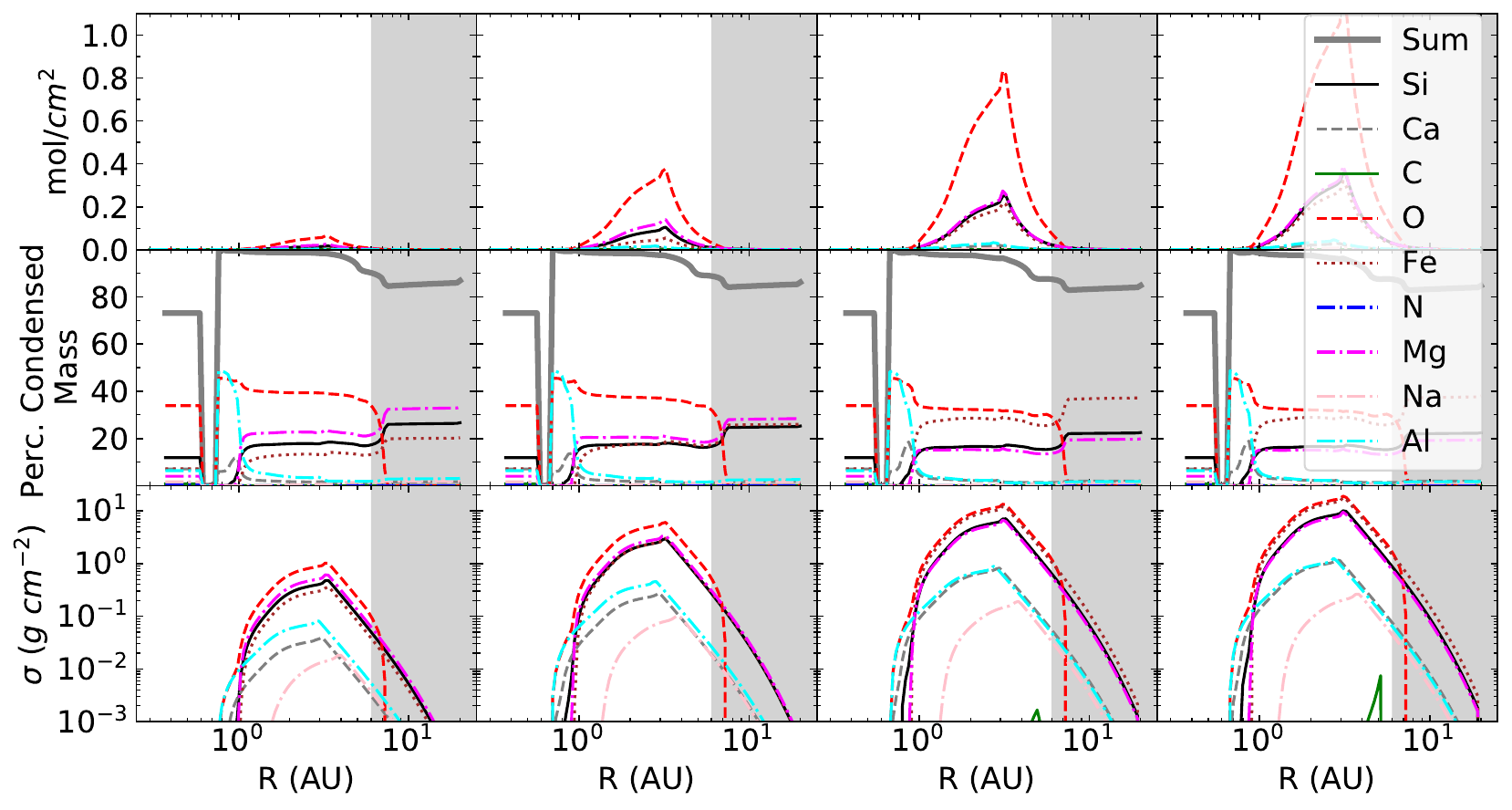}
    \caption{The final solid surface density of various elements in disks around 2 M$_\odot$ stars.  The columns and rows are the same as Figure \ref{fig:Disk1M}.}
    \label{fig:Disk2M}
\end{figure*}

Figures \ref{fig:Disk1M} and \ref{fig:Disk2M} show the surface density of condense materials for one and two-solar mass stars respectively. 
The grey areas of these figures are too cold for us to model rocky material condensation, as we only consider condensation temperatures above 300K.

The primary difference between the models is that the $2M_\odot$ star does not have significant condensation of material inside of $\sim0.5$AU, which may change with a more thorough examination of how stellar mass effects disk parameters.  With our model, the compositional difference caused by stars of different masses is small.

\subsection{Planet Properties}

Table \ref{tab:CMFs1sol} contains the core mass fraction (CMF) and mantle mass fraction (MMF) values used at each radius and time for the one and two solar mass simulations.  Table \ref{tab:CMFs1solHL} shows the CMF and MMF values taken from a disk enriched in elements from either high or low mass stars.

\begin{table*}[]
	\caption{Core mass fraction (CMF) and mantle mass fraction (MMF) values used to model planet interiors in MAGRATHEA for the one and two Solar mass simulation}
	\label{tab:CMFs1sol}
	\centering
        \scriptsize
        \tabcolsep=0.07cm
\begin{tabular}{cc|c|c|c|c|c|c|c|c|c|c|c|c|c}
\hline
\multicolumn{2}{c|}{\textbf{1 Solar Mass}} & \textbf{1 Gyr} & \textbf{2 Gyr} & \textbf{3 Gyr} & \textbf{4 Gyr} & \textbf{5 Gyr} & \textbf{6 Gyr} & \textbf{7 Gyr} & \textbf{8 Gyr} & \textbf{9 Gyr} & \textbf{10 Gyr} & \textbf{11 Gyr} & \textbf{12 Gyr} & \textbf{13 Gyr} \\ \hline
\multicolumn{1}{c|}{\multirow{2}{*}{0.5 AU}} & MMF & 1.00 & 0.975 & 0.919 & 0.86 & 0.8 & 0.747 & 0.761 & 0.759 & 0.757 & 0.756 & 0.755 & 0.754 & 0.753 \\
\multicolumn{1}{c|}{} & CMF & 0.00 & 0.025 & 0.081 & 0.14 & 0.2 & 0.253 & 0.239 & 0.241 & 0.243 & 0.244 & 0.245 & 0.246 & 0.247 \\ \hline
\multicolumn{1}{c|}{\multirow{2}{*}{1 AU}} & MMF & 0.863 & 0.867 & 0.818 & 0.771 & 0.723 & 0.694 & 0.702 & 0.701 & 0.7 & 0.699 & 0.699 & 0.698 & 0.698 \\
\multicolumn{1}{c|}{} & CMF & 0.137 & 0.133 & 0.182 & 0.229 & 0.277 & 0.306 & 0.298 & 0.299 & 0.3 & 0.301 & 0.301 & 0.302 & 0.302 \\ \hline
\multicolumn{1}{c|}{\multirow{2}{*}{2 AU}} & MMF & 0.86 & 0.864 & 0.815 & 0.769 & 0.723 & 0.695 & 0.703 & 0.702 & 0.701 & 0.7 & 0.7 & 0.699 & 0.699 \\
\multicolumn{1}{c|}{} & CMF & 0.14 & 0.136 & 0.185 & 0.231 & 0.277 & 0.305 & 0.297 & 0.298 & 0.299 & 0.3 & 0.3 & 0.301 & 0.301 \\ \hline
\multicolumn{1}{c|}{\multirow{2}{*}{3 AU}} & MMF & 0.859 & 0.862 & 0.813 & 0.768 & 0.722 & 0.695 & 0.703 & 0.702 & 0.701 & 0.7 & 0.699 & 0.698 & 0.698 \\
\multicolumn{1}{c|}{} & CMF & 0.141 & 0.138 & 0.187 & 0.232 & 0.278 & 0.305 & 0.297 & 0.298 & 0.299 & 0.3 & 0.301 & 0.302 & 0.302 \\ \hline
\multicolumn{1}{c|}{\multirow{2}{*}{4 AU}} & MMF & 0.852 & 0.853 & 0.804 & 0.761 & 0.72 & 0.696 & 0.703 & 0.702 & 0.701 & 0.7 & 0.699 & 0.698 & 0.698 \\
\multicolumn{1}{c|}{} & CMF & 0.148 & 0.147 & 0.196 & 0.239 & 0.28 & 0.304 & 0.297 & 0.298 & 0.299 & 0.3 & 0.301 & 0.302 & 0.302 \\ \hline
\end{tabular}

\begin{tabular}{cc|c|c|c|c|c|c|c|c|c|c|c|c|c}
\hline
\multicolumn{2}{c|}{\textbf{2 Solar Mass}} & \textbf{1 Gyr} & \textbf{2 Gyr} & \textbf{3 Gyr} & \textbf{4 Gyr} & \textbf{5 Gyr} & \textbf{6 Gyr} & \textbf{7 Gyr} & \textbf{8 Gyr} & \textbf{9 Gyr} & \textbf{10 Gyr} & \textbf{11 Gyr} & \textbf{12 Gyr} & \textbf{13 Gyr} \\ \hline
\multicolumn{1}{c|}{\multirow{2}{*}{1 AU}} & MMF & 1.00 & 0.984 & 0.927 & 0.866 & 0.797 & 0.743 & 0.76 & 0.758 & 0.756 & 0.754 & 0.753 & 0.752 & 0.751 \\
\multicolumn{1}{c|}{} & CMF & 0.00 & 0.016 & 0.073 & 0.134 & 0.203 & 0.257 & 0.24 & 0.242 & 0.244 & 0.246 & 0.247 & 0.248 & 0.249 \\ \hline
\multicolumn{1}{c|}{\multirow{2}{*}{2 AU}} & MMF & 0.86 & 0.865 & 0.815 & 0.767 & 0.719 & 0.689 & 0.697 & 0.696 & 0.695 & 0.695 & 0.694 & 0.693 & 0.693 \\
\multicolumn{1}{c|}{} & CMF & 0.14 & 0.135 & 0.185 & 0.233 & 0.281 & 0.311 & 0.303 & 0.304 & 0.305 & 0.305 & 0.306 & 0.307 & 0.307 \\ \hline
\multicolumn{1}{c|}{\multirow{2}{*}{3 AU}} & MMF & 0.86 & 0.863 & 0.814 & 0.768 & 0.722 & 0.693 & 0.702 & 0.7 & 0.699 & 0.698 & 0.698 & 0.697 & 0.697 \\
\multicolumn{1}{c|}{} & CMF & 0.14 & 0.137 & 0.186 & 0.232 & 0.278 & 0.307 & 0.298 & 0.3 & 0.301 & 0.302 & 0.302 & 0.303 & 0.303 \\ \hline
\multicolumn{1}{c|}{\multirow{2}{*}{4 AU}} & MMF & 0.853 & 0.853 & 0.805 & 0.762 & 0.722 & 0.699 & 0.705 & 0.704 & 0.703 & 0.702 & 0.702 & 0.701 & 0.701 \\
\multicolumn{1}{c|}{} & CMF & 0.147 & 0.147 & 0.195 & 0.238 & 0.278 & 0.301 & 0.295 & 0.296 & 0.297 & 0.298 & 0.298 & 0.299 & 0.299 \\ \hline
\end{tabular}
\end{table*}


\begin{table*}[]
\caption{CMF and MMF values for High and Low mass enriched simulations used to constrain the errors in planet composition at chosen ages for both the one and two solar mass cases.}
\label{tab:CMFs1solHL}
	\centering
        \scriptsize
        \tabcolsep=0.07cm
\begin{tabular}{cc|c|c|c|c|c|c|c|c}
\hline
\multicolumn{2}{c|}{\textbf{1 Solar Mass}} & \textbf{1 Gyr, Hi} & \textbf{1 Gyr, Lo} & \textbf{3 Gyr, Hi} & \textbf{3 Gyr, Lo} & \textbf{8 Gyr, Hi} & \textbf{8 Gyr, Lo} & \textbf{13 Gyr, Hi} & \textbf{13 Gyr, Lo} \\ \hline
\multicolumn{1}{c|}{\multirow{2}{*}{\textbf{0.5 AU}}} & MMF & 1.00 & 1.00 & 0.889 & 0.943 & 0.84 & 0.547 & 0.883 & 0.508 \\
\multicolumn{1}{c|}{} & CMF & 0.00 & 0.00 & 0.057 & 0.111 & 0.16 & 0.453 & 0.167 & 0.492 \\ \hline
\multicolumn{1}{c|}{\multirow{2}{*}{\textbf{1 AU}}} & MMF & 0.889 & 0.863 & 0.856 & 0.788 & 0.762 & 0.603 & 0.759 & 0.579 \\
\multicolumn{1}{c|}{} & CMF & 0.111 & 0.137 & 0.144 & 0.212 & 0.238 & 0.397 & 0.241 & 0.421 \\ \hline
\multicolumn{1}{c|}{\multirow{2}{*}{\textbf{2 AU}}} & MMF & 0.886 & 0.86 & 0.852 & 0.786 & 0.76 & 0.618 & 0.757 & 0.606 \\
\multicolumn{1}{c|}{} & CMF & 0.114 & 0.14 & 0.148 & 0.214 & 0.24 & 0.382 & 0.243 & 0.394 \\ \hline
\multicolumn{1}{c|}{\multirow{2}{*}{\textbf{3 AU}}} & MMF & 0.884 & 0.859 & 0.849 & 0.784 & 0.757 & 0.622 & 0.754 & 0.61 \\
\multicolumn{1}{c|}{} & CMF & 0.116 & 0.141 & 0.151 & 0.216 & 0.243 & 0.378 & 0.246 & 0.39 \\ \hline
\multicolumn{1}{c|}{\multirow{2}{*}{\textbf{4 AU}}} & MMF & 0.876 & 0.852 & 0.838 & 0.777 & 0.749 & 0.641 & 0.745 & 0.635 \\
\multicolumn{1}{c|}{} & CMF & 0.124 & 0.148 & 0.162 & 0.223 & 0.251 & 0.359 & 0.255 & 0.365 \\ \hline
\end{tabular}

\begin{tabular}{cc|c|c|c|c|c|c|c|c}
\hline
\multicolumn{2}{c|}{\textbf{2 Solar Mass}} & \textbf{1 Gyr, Hi} & \textbf{1 Gyr, Lo} & \textbf{3 Gyr, Hi} & \textbf{3 Gyr, Lo} & \textbf{8 Gyr, Hi} & \textbf{8 Gyr, Lo} & \textbf{13 Gyr, Hi} & \textbf{13 Gyr, Lo} \\ \hline
\multicolumn{1}{c|}{\multirow{2}{*}{\textbf{1 AU}}} & MMF & 1.00 & 1.00 & 0.95 & 0.897 & 0.845 & 0.469 & 0.837 & 0.416 \\
\multicolumn{1}{c|}{} & CMF & 0.00 & 0.00 & 0.05 & 0.103 & 0.155 & 0.531 & 0.163 & 0.584 \\ \hline
\multicolumn{1}{c|}{\multirow{2}{*}{\textbf{2 AU}}} & MMF & 0.887 & 0.86 & 0.853 & 0.784 & 0.757 & 0.605 & 0.758 & 0.589 \\
\multicolumn{1}{c|}{} & CMF & 0.113 & 0.14 & 0.147 & 0.216 & 0.243 & 0.395 & 0.245 & 0.411 \\ \hline
\multicolumn{1}{c|}{\multirow{2}{*}{\textbf{3 AU}}} & MMF & 0.885 & 0.86 & 0.852 & 0.785 & 0.76 & 0.615 & 0.758 & 0.6 \\
\multicolumn{1}{c|}{} & CMF & 0.115 & 0.14 & 0.148 & 0.215 & 0.24 & 0.385 & 0.242 & 0.4 \\ \hline
\multicolumn{1}{c|}{\multirow{2}{*}{\textbf{4 AU}}} & MMF & 0.876 & 0.853 & 0.838 & 0.778 & 0.75 & 0.646 & 0.747 & 0.644 \\
\multicolumn{1}{c|}{} & CMF & 0.124 & 0.147 & 0.162 & 0.222 & 0.25 & 0.354 & 0.253 & 0.356 \\ \hline
\end{tabular}
\end{table*}

\bibliography{ageplanet}{}

\begin{thebibliography}{}
\expandafter\ifx\csname natexlab\endcsname\relax\def\natexlab#1{#1}\fi
\providecommand{\url}[1]{\href{#1}{#1}}
\providecommand{\dodoi}[1]{doi:~\href{http://doi.org/#1}{\nolinkurl{#1}}}
\providecommand{\doeprint}[1]{\href{http://ascl.net/#1}{\nolinkurl{http://ascl.net/#1}}}
\providecommand{\doarXiv}[1]{\href{https://arxiv.org/abs/#1}{\nolinkurl{https://arxiv.org/abs/#1}}}

\bibitem[{Adibekyan {et~al.}(2021)Adibekyan, Dorn, Sousa, Santos, Bitsch,
  Israelian, Mordasini, Barros, Mena, Demangeon, Faria, Figueira, Hakobyan,
  Oshagh, Soares, Kunitomo, Takeda, Jofré, Petrucci, \&
  Martioli}]{Adibekyan2021}
Adibekyan, V., Dorn, C., Sousa, S.~G., {et~al.} 2021, Science, 374, 330,
  \dodoi{10.1126/science.abg8794}

\bibitem[{{Bogdan} {et~al.}(2023){Bogdan}, {Pillich}, {Landers}, {Wende}, \&
  {Wurm}}]{Bogdan2023}
{Bogdan}, T., {Pillich}, C., {Landers}, J., {Wende}, H., \& {Wurm}, G. 2023,
  \aap, 670, A6, \dodoi{10.1051/0004-6361/202245106}

\bibitem[{{Boley} {et~al.}(2024){Boley}, {Christiansen}, {Zink},
  {Hardegree-Ullman}, {Lee}, {Hopkins}, {Wang}, {Fernandes}, {Bergsten}, \&
  {Bhure}}]{Boley:2024}
{Boley}, K.~M., {Christiansen}, J.~L., {Zink}, J., {et~al.} 2024, \aj, 168,
  128, \dodoi{10.3847/1538-3881/ad6570}

\bibitem[{{Bond} {et~al.}(2010){Bond}, {O'Brien}, \& {Lauretta}}]{Bond2010}
{Bond}, J.~C., {O'Brien}, D.~P., \& {Lauretta}, D.~S. 2010, \apj, 715, 1050,
  \dodoi{10.1088/0004-637X/715/2/1050}

\bibitem[{{Brewer} \& {Fischer}(2016)}]{Brewers2016}
{Brewer}, J.~M., \& {Fischer}, D.~A. 2016, \apj, 831, 20,
  \dodoi{10.3847/0004-637X/831/1/20}

\bibitem[{{Brinkman} {et~al.}(2024){Brinkman}, {Polanski}, {Huber}, {Weiss},
  {Valencia}, \& {Plotnykov}}]{Brinkman2024}
{Brinkman}, C.~L., {Polanski}, A.~S., {Huber}, D., {et~al.} 2024, \aj, 168,
  281, \dodoi{10.3847/1538-3881/ad82eb}

\bibitem[{{Cabral} {et~al.}(2023){Cabral}, {Guilbert-Lepoutre}, {Bitsch},
  {Lagarde}, \& {Diakite}}]{Cabral:2023}
{Cabral}, N., {Guilbert-Lepoutre}, A., {Bitsch}, B., {Lagarde}, N., \&
  {Diakite}, S. 2023, \aap, 673, A117, \dodoi{10.1051/0004-6361/202243882}

\bibitem[{{Cassen}(1996)}]{Cassen1996}
{Cassen}, P. 1996, \maps, 31, 793, \dodoi{10.1111/j.1945-5100.1996.tb02114.x}

\bibitem[{{Ferich} {et~al.}(2025){Ferich}, {Childs}, \&
  {Steffen}}]{Ferich:2025}
{Ferich}, N., {Childs}, A.~C., \& {Steffen}, J.~H. 2025, \na, 114, 102315,
  \dodoi{10.1016/j.newast.2024.102315}

\bibitem[{Goldschmidt(1937)}]{goldschmidt}
Goldschmidt, V.~M. 1937, J. Chem. Soc., 655, \dodoi{10.1039/JR9370000655}

\bibitem[{Harris {et~al.}(2020)Harris, Millman, van~der Walt, Gommers,
  Virtanen, Cournapeau, Wieser, Taylor, Berg, Smith, Kern, Picus, Hoyer, van
  Kerkwijk, Brett, Haldane, del R{\'{i}}o, Wiebe, Peterson,
  G{\'{e}}rard-Marchant, Sheppard, Reddy, Weckesser, Abbasi, Gohlke, \&
  Oliphant}]{numpy}
Harris, C.~R., Millman, K.~J., van~der Walt, S.~J., {et~al.} 2020, Nature, 585,
  357, \dodoi{10.1038/s41586-020-2649-2}

\bibitem[{{Haywood} {et~al.}(2016){Haywood}, {Lehnert}, {Di Matteo}, {Snaith},
  {Schultheis}, {Katz}, \& {G{\'o}mez}}]{Haywood:2016}
{Haywood}, M., {Lehnert}, M.~D., {Di Matteo}, P., {et~al.} 2016, \aap, 589,
  A66, \dodoi{10.1051/0004-6361/201527567}

\bibitem[{{Huang} {et~al.}(2022){Huang}, {Rice}, \& {Steffen}}]{Magrathea}
{Huang}, C., {Rice}, D.~R., \& {Steffen}, J.~H. 2022, \mnras, 513, 5256,
  \dodoi{10.1093/mnras/stac1133}

\bibitem[{Hunter(2007)}]{matplotlib}
Hunter, J.~D. 2007, Computing in Science \& Engineering, 9, 90,
  \dodoi{10.1109/MCSE.2007.55}

\bibitem[{{Jacquet} {et~al.}(2012){Jacquet}, {Gounelle}, \&
  {Fromang}}]{Jacquet2012}
{Jacquet}, E., {Gounelle}, M., \& {Fromang}, S. 2012, \icarus, 220, 162,
  \dodoi{10.1016/j.icarus.2012.04.022}

\bibitem[{{Johnson}(2019)}]{Johnson2019}
{Johnson}, J.~A. 2019, Science, 363, 474, \dodoi{10.1126/science.aau9540}

\bibitem[{{Kobayashi} {et~al.}(2020){Kobayashi}, {Karakas}, \&
  {Lugaro}}]{Kobayashi2020}
{Kobayashi}, C., {Karakas}, A.~I., \& {Lugaro}, M. 2020, \apj, 900, 179,
  \dodoi{10.3847/1538-4357/abae65}

\bibitem[{{Kroupa} {et~al.}(2024){Kroupa}, {Gjergo}, {Jerabkova}, \&
  {Yan}}]{Kroupa:2024}
{Kroupa}, P., {Gjergo}, E., {Jerabkova}, T., \& {Yan}, Z. 2024, arXiv e-prints,
  arXiv:2410.07311, \dodoi{10.48550/arXiv.2410.07311}

\bibitem[{{Li} {et~al.}(2020){Li}, {Huang}, {Petaev}, {Zhu}, \&
  {Steffen}}]{Li2020}
{Li}, M., {Huang}, S., {Petaev}, M.~I., {Zhu}, Z., \& {Steffen}, J.~H. 2020,
  \mnras, 495, 2543, \dodoi{10.1093/mnras/staa1149}

\bibitem[{{Lodders}(2003)}]{Lodders2003}
{Lodders}, K. 2003, \apj, 591, 1220, \dodoi{10.1086/375492}

\bibitem[{{Luo} {et~al.}(2024){Luo}, {Dorn}, \& {Deng}}]{Luo:2024}
{Luo}, H., {Dorn}, C., \& {Deng}, J. 2024, Nature Astronomy, 8, 1399,
  \dodoi{10.1038/s41550-024-02347-z}

\bibitem[{{Maoz} \& {Graur}(2017)}]{Maoz:2017}
{Maoz}, D., \& {Graur}, O. 2017, \apj, 848, 25,
  \dodoi{10.3847/1538-4357/aa8b6e}

\bibitem[{{Marcus} {et~al.}(2010){Marcus}, {Sasselov}, {Hernquist}, \&
  {Stewart}}]{Marcus2010}
{Marcus}, R.~A., {Sasselov}, D., {Hernquist}, L., \& {Stewart}, S.~T. 2010,
  \apjl, 712, L73, \dodoi{10.1088/2041-8205/712/1/L73}

\bibitem[{Ness {et~al.}(2019)Ness, Johnston, Blancato, Rix, Beane, Bird, \&
  Hawkins}]{Ness:2019}
Ness, M.~K., Johnston, K.~V., Blancato, K., {et~al.} 2019, The Astrophysical
  Journal, 883, 177, \dodoi{10.3847/1538-4357/ab3e3c}

\bibitem[{{Nissen}(2015)}]{Nissen:2015}
{Nissen}, P.~E. 2015, \aap, 579, A52, \dodoi{10.1051/0004-6361/201526269}

\bibitem[{Petaev(2009)}]{Petaev2009}
Petaev, M.~I. 2009, Calphad, 33, 317, \dodoi{10.1016/j.calphad.2008.12.001}

\bibitem[{{Rauer} {et~al.}(2025){Rauer}, {Aerts}, {Cabrera}, {Deleuil},
  {Erikson}, {Gizon}, {Goupil}, {Heras}, {Walloschek}, {Lorenzo-Alvarez},
  {Marliani}, {Martin-Garcia}, {Mas-Hesse}, {O'Rourke}, {Osborn}, {Pagano},
  {Piotto}, {Pollacco}, {Ragazzoni}, {Ramsay}, {Udry}, {Appourchaux}, {Benz},
  {Brandeker}, {G{\"u}del}, {Janot-Pacheco}, {Kabath}, {Kjeldsen}, {Min},
  {Santos}, {Smith}, {Suarez}, {Werner}, {Aboudan}, {Abreu}, {Acu{\~n}a},
  {Adams}, {Adibekyan}, {Affer}, {Agneray}, {Agnor}, {Aguirre B{\o}rsen-Koch},
  {Ahmed}, {Aigrain}, {Al-Bahlawan}, {Alcacera Gil}, {Alei}, {Alencar},
  {Alexander}, {Alfonso-Garz{\'o}n}, {Alibert}, {Allende Prieto}, {Almeida},
  {Alonso Sobrino}, {Altavilla}, {Althaus}, {Alvarez Trujillo}, {Amarsi},
  {Ammler-von Eiff}, {Am{\^o}res}, {Andrade}, {Antoniadis-Karnavas},
  {Ant{\'o}nio}, {Aparicio del Moral}, {Appolloni}, {Arena}, {Armstrong},
  {Aroca Aliaga}, {Asplund}, {Audenaert}, {Auricchio}, {Avelino}, {Baeke},
  {Bailli{\'e}}, {Balado}, {Ballber Balaguer{\'o}}, {Balestra}, {Ball},
  {Ballans}, {Ballot}, {Barban}, {Barbary}, {Barbieri}, {Barcel{\'o} Forteza},
  {Barker}, {Barklem}, {Barnes}, {Barrado Navascues}, {Barragan}, {Baruteau},
  {Basu}, {Baudin}, {Baumeister}, {Bayliss}, {Bazot}, {Beck}, {Belkacem},
  {Bellinger}, {Benatti}, {Benomar}, {B{\'e}rard}, {Bergemann}, {Bergomi},
  {Bernardo}, {Biazzo}, {Bignamini}, {Bigot}, {Billot}, {Binet}, {Biondi},
  {Biondi}, {Birch}, {Bitsch}, {Bluhm Ceballos}, {B{\'o}di}, {Bogn{\'a}r},
  {Boisse}, {Bolmont}, {Bonanno}, {Bonavita}, {Bonfanti}, {Bonfils}, {Bonito},
  {Bonomo}, {B{\"o}rner}, {Boro Saikia}, {Borreguero Mart{\'\i}n}, {Borsa},
  {Borsato}, {Bossini}, {Bouchy}, {Bou{\'e}}, {Boufleur}, {Boumier},
  {Bourrier}, {Bowman}, {Bozzo}, {Bradley}, {Bray}, {Bressan}, {Breton},
  {Brienza}, {Brito}, {Brogi}, {Brown}, {Brown}, {Brun}, {Bruno}, {Bruns},
  {Buchhave}, {Bugnet}, {Buldgen}, {Burgess}, {Busatta}, {Busso}, {Buzasi},
  {Caballero}, {Cabral}, {Cabrero Gomez}, {Calderone}, {Cameron}, {Cameron},
  {Campante}, {Campos Gestal}, {Canto Martins}, {Cara}, {Carone}, {Carrasco},
  {Casagrande}, {Casewell}, {Cassisi}, {Castellani}, {Castro}, {Catala},
  {Catal{\'a}n Fern{\'a}ndez}, {Catelan}, {Cegla}, {Cerruti}, {Cessa},
  {Chadid}, {Chaplin}, {Charpinet}, {Chiappini}, {Chiarucci}, {Chiavassa},
  {Chinellato}, {Chirulli}, {Christensen-Dalsgaard}, {Church}, {Claret},
  {Clarke}, {Claudi}, {Clermont}, {Coelho}, {Coelho}, {Cogato}, {Colom{\'e}},
  {Condamin}, {Conde Garc{\'\i}a}, \& {Conseil}}]{Rauer:2025}
{Rauer}, H., {Aerts}, C., {Cabrera}, J., {et~al.} 2025, Experimental Astronomy,
  59, 26, \dodoi{10.1007/s10686-025-09985-9}

\bibitem[{{Rice} {et~al.}(2025){Rice}, {Huang}, {Steffen}, \&
  {Vazan}}]{Rice2025}
{Rice}, D.~R., {Huang}, C., {Steffen}, J.~H., \& {Vazan}, A. 2025, arXiv
  e-prints, arXiv:2504.16201, \dodoi{10.48550/arXiv.2504.16201}

\bibitem[{{Rogers} {et~al.}(2025){Rogers}, {Dorn}, {Aditya Raj}, {Schlichting},
  \& {Young}}]{Rogers:2025}
{Rogers}, J.~G., {Dorn}, C., {Aditya Raj}, V., {Schlichting}, H.~E., \&
  {Young}, E.~D. 2025, \apj, 979, 79, \dodoi{10.3847/1538-4357/ad9f61}

\bibitem[{{Salaris} \& {Cassisi}(2006)}]{Salaris:2006}
{Salaris}, M., \& {Cassisi}, S. 2006, {Evolution of Stars and Stellar
  Populations}

\bibitem[{{Salpeter}(1955)}]{Salpeter:1955}
{Salpeter}, E.~E. 1955, \apj, 121, 161, \dodoi{10.1086/145971}

\bibitem[{{Shakespeare} {et~al.}(2025){Shakespeare}, {Li}, {Huang}, {Zhu}, \&
  {Steffen}}]{Shake2025}
{Shakespeare}, C.~J., {Li}, M., {Huang}, S., {Zhu}, Z., \& {Steffen}, J.~H.
  2025, \aj, 169, 180, \dodoi{10.3847/1538-3881/adaead}

\bibitem[{{Snaith} {et~al.}(2015){Snaith}, {Haywood}, {Di Matteo}, {Lehnert},
  {Combes}, {Katz}, \& {G{\'o}mez}}]{Snaith:2015}
{Snaith}, O., {Haywood}, M., {Di Matteo}, P., {et~al.} 2015, \aap, 578, A87,
  \dodoi{10.1051/0004-6361/201424281}

\bibitem[{{Teixeira} {et~al.}(2025){Teixeira}, {Adibekyan}, \&
  {Bossini}}]{Teixeira:2025}
{Teixeira}, J., {Adibekyan}, V., \& {Bossini}, D. 2025, Astronomische
  Nachrichten, 346, e20240076, \dodoi{10.1002/asna.20240076}

\bibitem[{{Tucci Maia} {et~al.}(2016){Tucci Maia}, {Ram{\'\i}rez},
  {Mel{\'e}ndez}, {Bedell}, {Bean}, \& {Asplund}}]{TucciMaia:2016}
{Tucci Maia}, M., {Ram{\'\i}rez}, I., {Mel{\'e}ndez}, J., {et~al.} 2016, \aap,
  590, A32, \dodoi{10.1051/0004-6361/201527848}

\bibitem[{{Weeks} {et~al.}(2025){Weeks}, {Van Eylen}, {Huber}, {Kawata},
  {Stokholm}, {B{\o}rsen-Koch}, {Pinilla}, {R{\o}rsted}, {Winther}, \&
  {Berger}}]{Weeks2025}
{Weeks}, A., {Van Eylen}, V., {Huber}, D., {et~al.} 2025, \mnras,
  \dodoi{10.1093/mnras/staf474}

\end{thebibliography}
\bibliographystyle{aasjournal}



\end{document}